\documentclass[twocolumn,english,prl,floatfix,citeautoscript,nofootinbib]{revtex4}
\usepackage{amsbsy}
\usepackage{latexsym,epsfig,graphicx}
\usepackage{dcolumn}
\usepackage{subfigure}
\usepackage{comment}
\usepackage{color}
\usepackage[colorlinks,urlcolor=blue,citecolor=blue]{hyperref}
\usepackage{amstext}
\usepackage{amssymb}
\usepackage{setspace}
\usepackage{amsmath}
\usepackage{makeidx}
\usepackage{bm}

\DeclareMathOperator{\sech}{sech}

\begin{document}

\title{Magnetic stripe soliton and localized stripe wave in spin-1
Bose-Einstein condensates}
\author{Li-Chen Zhao$^{1,2}$}
\author{Xi-Wang Luo$^1$}
\author{Chuanwei Zhang$^1$}
\email{chuanwei.zhang@utdallas.edu}
\address{$^{1}$Department of Physics, The University of Texas at Dallas,
Richardson, Texas 75080, USA}
\address{$^{2}$School of Physics, Northwest
University, Xi'an, 710069, China}

\begin{abstract}
The recent experimental realization of spin-orbit coupling for ultracold
atomic gases opens a new avenue for engineering solitons with internal
spatial structures through tuning atomic band dispersions. However, the
types of the resulting stripe solitons in a spin-1/2 Bose-Einstein
condensate (BEC) have been limited to dark-dark or bright-bright with the
same density profiles for different spins. Here we propose that general
types of stripe solitons, including magnetic stripe (e.g., dark-bright) and
localized stripe waves (neither bright nor dark), could be realized in a
spin-1 BEC with widely tunable band dispersions through modulating the
coupling between three spin states and the linear momentum of atoms. \
Surprisingly, a moving magnetic stripe soliton can possess both negative and
positive effective masses at different velocities, leading to a zero mass
soliton at certain velocity. Our work showcases the great potential of
realizing novel types of solitons through band dispersion engineering, which
may provide a new approach for exploring soliton physics in many physical
branches.
\end{abstract}

\maketitle


\emph{Introduction}: Solitons are topological defects that play significant
roles in many different physical branches, such as water waves, fiber
optics, and nonlinear matter waves \cite{S2,Drazin,Dickey,Kivshar,S6}.
Ultra-cold atomic superfluids, such as Bose-Einstein condensates (BECs) and
degenerate Fermi gases (DFGs), provide a disorder-free and highly
controllable platform for exploring soliton physics \cite%
{Denschlag,Anderson,fermib,Meyer,S5,S7,BS,S8,S4,BEC,Ebd,Bersano}. In
particular, because solitons arise from the interplay between band
dispersion and nonlinearity, tuning atomic interactions through Feshbach
resonance~\cite{RevModPhys.82.1225} provides a tunable knob for generating
various types of solitons and exploring their dynamical properties. For
instance, a dark (bright) soliton has been observed in a single component
BEC with repulsive (attractive) interaction \cite{S5,S7,BS}, while their
combinations, such as dark-dark and dark-bright solitons have been realized
in multiple-component BECs with different inter- and intra-component
interactions \cite{S4,BEC,Ebd}. Notably, a \textquotedblleft magnetic
soliton" with a uniform total atom density has been predicted recently for a
two-component BEC \cite{Qu1,Qu2}, where the dark soliton of one component is
perfectly filled by the anti-dark soliton of the other component.

The recent experimental realization of spin-orbit coupling for ultracold
atoms~\cite{lin2011spin, zhang2012collective, qu2013observation,
olson2014tunable, wang2012spin, Cheuk2012, Williams2013,
huang2016experimental, wu2016realization, campbell2015itinerant,
luo2016tunable} opens a new avenue for exploring soliton physics through
engineering atomic band dispersions~\cite%
{Fialko,Zhou,Kartashov,stripeso,Wu,Xu,Achilleos,stripeso2}. In a spin-1/2
BEC, the spin-orbit coupling displays a double well band dispersion, and the
simultaneous occupation of two momentum space minima opens the possibility
for generating solitons with internal spatial structures, i.e., the stripe
density modulation \cite{stripeso,Wu,Xu,Achilleos,stripeso2}. However, the
intrinsic Raman coupling for the realization of spin-orbit coupling mixes
two spin states, which demands the same spatial density for different spins,
therefore only certain types of stripe solitons such as bright-bright or
dark-dark can exist \cite{stripeso,Wu,Xu,Achilleos,stripeso2}. A natural
question is whether general types of stripe solitons, such as magnetic
stripe solitons (dark-bright or dark-antidark types) could be generated by
tuning the band dispersion beyond that for spin-1/2 spin-orbit coupling.

In this paper, we address this important question by considering a spin-1
BEC with widely tunable band dispersion achieved by coupling three spin
states $\left\{ \left\vert \uparrow \right\rangle ,\left\vert 0\right\rangle
,\left\vert \downarrow \right\rangle \right\} $ with the linear momentum of
atoms. Our main results are:

\emph{i}) Both dark-bright and dark-antidark stripe solitons could exist for
ferromagnetic or antiferromagnetic spin interactions. The dark and bright
solitons reside at two band minima with different momenta and the spin
states $\left\vert \uparrow \right\rangle $ and $\left\vert \downarrow
\right\rangle $ exhibit strong stripe density modulations on top of a
soliton background.

\emph{ii}) By slightly tuning the spin interactions, a magnetic stripe
soliton with a uniform total density could be generated. The dark-soliton
dip in the state $|0\rangle $ is perfectly filled by the bright-soliton
atoms in states $\left\vert \uparrow \right\rangle $ and $\left\vert
\downarrow \right\rangle $, which exhibit an out-of-phase density
modulation. More interestingly, the soliton's effective mass can possess
both positive and negative values at different velocities, in contrast to
solitons with a fixed sign of mass in previous literature \cite%
{Qu1,eff1,eff2,eff3,eff4,Scott}.

\emph{iii}) When the spin interaction is comparable with the density
interaction, a localized stripe wave for states $\left\vert \uparrow
\right\rangle $ and $\left\vert \downarrow \right\rangle $, which is neither
bright nor dark, could exist. Two local stripe density modulations reside on
the same uniform background, but cancel each other, leading to a uniform
spin tensor density.

\emph{iv}) Such magnetic stripe solitons and localized stripe waves are
stable and their stability is confirmed by numerically simulating the
mean-field dynamical equations.

\begin{figure}[h]
\begin{center}
\includegraphics[width=1.0\linewidth]{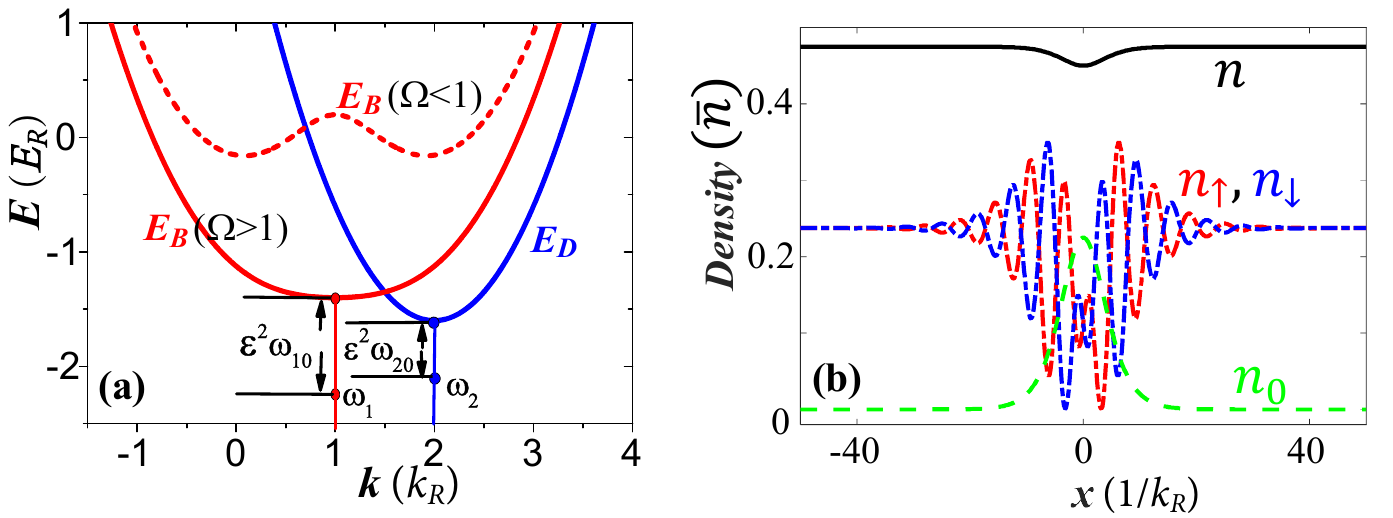}
\end{center}
\caption{(a) Single-particle band structures with detuning $\Delta =-\protect%
\delta =-0.8$. The red solid (dashed) line denotes the bright band for $%
\Omega =1.2$ ($\Omega =0.4$), and the blue line denotes the dark band
(independent of $\Omega $). The $\protect\epsilon ^{2}\protect\omega _{j0}$ $%
(j=1,2)$ corresponds to the nonlinear energy correction induced by the
soliton. (b) The spin-density distributions for a typical stripe soliton
solution with anti-ferromagnetic interaction $g_{2}>0$. $n$ is the total
density. The parameters are $\protect\epsilon =0.1$, $g_{2}=-0.1$, $g_{0}=1$%
, $\protect\delta =0.5$, $\protect\omega _{10}=-42.75$, $\protect\omega %
_{20}=-47.5$.}
\label{Fig1}
\end{figure}

\emph{System and methods}:\emph{\ }We consider the emergence of stripe
solitons on two spin states $\left\vert \uparrow \right\rangle $ and $%
\left\vert \downarrow \right\rangle $ through their coupling with the third
state $\left\vert 0\right\rangle $ in a spin-1 BEC. To avoid large momentum
transfer between $\left\vert \uparrow \right\rangle $ and $\left\vert
\downarrow \right\rangle $, while still modify their local band dispersions,
we adopt the recently proposed spin-tensor-momentum coupling for a spin-1
system \cite{Luo,Luo2}, which can be realized by coupling three hyperfine
ground states (denoted as $\{\left\vert \uparrow \right\rangle ,\left\vert
0\right\rangle ,\left\vert \downarrow \right\rangle \}$) of Alkali atoms
through Raman and microwave transitions [similar scheme also applies to
alkali-earth(-like) atoms]. The single-particle Hamiltonian under the basis $%
\{\left\vert \uparrow \right\rangle ,\left\vert 0\right\rangle ,\left\vert
\downarrow \right\rangle \}$ reads
\begin{equation}
H_{0}=(i\partial _{x}+2F_{z}^{2})^{2}+\Delta F_{z}^{2}+\sqrt{2}\Omega
F_{x}+\delta (\left\vert \uparrow \right\rangle \left\langle \downarrow
\right\vert +h.c.).
\end{equation}%
Here we set $\hbar =1$ and the recoil energy $E_R=k_{R}^{2}/2m$ and momentum $%
k_{R}$ as energy and momentum units. $F_{z}$ and $F_{x}$ are spin-1 vectors,
$\Delta $ is the tensor Zeeman field, $\Omega $ is the Raman coupling
strength between states $\left\vert \uparrow ,\downarrow \right\rangle $ and
$\left\vert 0\right\rangle $, and $\delta $ describes the microwave coupling
between $\left\vert \uparrow \right\rangle $ and $\left\vert \downarrow
\right\rangle $. The typical lower band structure is shown in Fig.~\ref{Fig1}%
(a), with the low energy dynamics characterized by two bands: the dark band $%
E_{D}(k)$ (blue line) for the state $\left\vert -\right\rangle $; the bright
band $E_{B}(k)$ (red lines) for the mixture of the states $\left\vert
+\right\rangle $ and $\left\vert 0\right\rangle $, where $\left\vert \pm
\right\rangle =\frac{1}{\sqrt{2}}(\left\vert \uparrow \right\rangle \pm
\left\vert \downarrow \right\rangle )$). The dark band has a minimum at $k=2$%
, while the bright band may have one or two band minima depending on the
Raman coupling strength $\Omega $.

The interaction between atoms, which is needed for generating solitons, can
be described under the mean field approximation. For the simplicity of the
calculation, the spin basis $\{\left\vert +\right\rangle ,\left\vert
0\right\rangle ,\left\vert -\right\rangle \}$ is used with the corresponding
mean-field Gross-Pitaevskii equation (see Appendix)
\begin{equation}
i\partial _{t}\psi _{j}=H_{0}\psi _{j}+(g_{0}\bar{n}+g_{2}\bar{n})n\psi
_{j}-g_{2}|\psi _{j}|^{2}\psi _{j}+\delta _{j,0}\gamma |\psi _{j}|^{2}\psi _{j},  \label{eq:eom}
\end{equation}%
where $\psi _{j}$ is the wavefunction at the spin state $|j\rangle $, $g_{2}$
and $g_{0}>|g_{2}|$ are the spin and density interaction strengths,
respectively, and $n$ is the atom density with density unit $\bar{n}$ (we
set $\bar{n}=1$ without loss of generality). The term $%
\delta _{j,0}\gamma |\psi _{j}|^{2}\psi _{j}$ in the right-hand side of Eq.~%
\ref{eq:eom} corresponds to additional intra-component
interaction $\frac{\gamma }{2}|\psi _{0}|^{4}$ for spin state $\left\vert
0\right\rangle $, which can be realized using Feshbach resonance~\cite{RevModPhys.82.1225}.
Hereafter $\Delta =-\delta $ is
chosen so that the bright band is symmetric around $k=1$ to obtain simple
analytic soliton solutions.

Since exact analytic soliton solutions cannot be obtained for such a spin-1
system, here we derive an approximate solution using the multi-scale
expansion method~\cite{stripeso,stripeso2}. We choose an ansatz for the
wavefunctions of the soliton%
\begin{equation}
\psi _{+(0)}=A_{+(0)}\left( x\right) e^{ik_{1}x-i\omega _{1}t},\text{ }\psi
_{-}=A_{-}\left( x\right) e^{ik_{2}x-i\omega _{2}t},  \label{eq:ansatz}
\end{equation}%
where $A_{+}$, $A_{0}$ and $A_{-}$ (with $|A_{j}|\ll 1$) describe the
spatial profiles of the soliton, $k_{1}$, $k_{2}$ are the center momenta of
the soliton, and $\omega _{1}$, $\omega _{2}$ are the soliton energies. The
soliton amplitudes can be expanded as $A_{j}=\sum_{\eta }\epsilon ^{\eta
+1}\chi _{j}^{(\eta )}$ using a small parameter $\epsilon $, where $\chi
_{j}^{(\eta )}$ are slowly varying functions (i.e. $\partial _{x}\chi
_{j}^{(\eta )}\sim \epsilon \chi _{j}^{(\eta )}$). Substituting the ansatz (%
\ref{eq:ansatz}) into Eq.~(\ref{eq:eom}), we obtain the soliton solution and
corresponding constrains on $\chi _{j}^{(\eta )}$ according to the
solvability conditions up to the order of $O(\epsilon ^{3})$. It is easy to
check that the leading order solution satisfies $\chi _{+}^{(0)}=-\chi
_{0}^{(0)}\equiv U(X)$ and $\chi _{-}^{(0)}\equiv V(X)$ with $X=\epsilon x$.

\emph{Dark-bright stripe soliton}: Here we focus on the case $\Omega =1$
where the bright band has a minimum at $k=1$ and the dispersion effects are
suppressed significantly. The soliton solution is static with $k_{1}=1$, $%
k_{2}=2$. The leading order in the wavefunction (\ref{eq:ansatz}) yields $%
A_{+}\left( x\right) =-A_{0}\left( x\right) \approx \epsilon U(X)$, $%
A_{-}\left( x\right) =\epsilon V(X)$, $\omega _{1}\approx E_{B}(1)-\epsilon
^{2}\omega _{10}$, and $\omega _{2}\approx E_{D}(2)-\epsilon ^{2}\omega
_{20} $, with the energy corrections $\epsilon ^{2}\omega _{10}$ and $%
\epsilon ^{2}\omega _{20}$ induced by the nonlinear interaction. For $\gamma=0$, the
mean-field equation (\ref{eq:eom}) can be approximated as
\begin{equation}
\partial _{X}^{2}V(X)+g_{V}|V(X)|^{2}V(X)+w_{V}V(X)=0,  \label{eq:eom2}
\end{equation}%
where the coefficients $g_{V}=\left( 2g_{2}^{2}+3g_{0}g_{2}\right) /\left(
2g_{0}+g_{2}\right) $ and $w_{V}=2(g_{0}+g_{2})\omega _{10}/\left(
2g_{0}+g_{2}\right) -\omega _{20}$. $U(X)$ can be determined through the
constrain
\begin{equation}
|U(X)|^{2}+\frac{g_{0}+g_{2}}{2g_{0}+g_{2}}|V(X)|^{2}=\frac{-\omega _{10}}{%
2g_{0}+g_{2}}>0.  \label{eq:uv}
\end{equation}

Since $g_{V}$ has the same sign as $g_{2}$, the effective interaction
strength $g_{V}<0$ or $g_{V}>0$ in Eq. (\ref{eq:eom2}) for ferromagnetic ($%
g_{2}<0$) and antiferromagnetic ($g_{2}>0$) spin interactions, respectively.
For $g_{V}<0$, Eqs. (\ref{eq:eom2}) and (\ref{eq:uv}) have a dark soliton
solution
\begin{equation}
V(X)=\sqrt{\frac{w_{V}}{-g_{V}}}\tanh [\sqrt{\frac{w_{V}}{2}}X]
\label{eq:vform1}
\end{equation}%
in the parameter region $\frac{g_{0}+g_{2}}{g_{V}}\geq \frac{\omega _{10}}{%
\omega _{V}}$ and $w_{V}>0$. The corresponding $U$ from Eq. (\ref{eq:uv}) is
a bright or anti-dark soliton. While for $g_{V}>0$,
\begin{equation}
V(X)=\sqrt{\frac{-2w_{V}}{g_{V}}}\sech[\sqrt{- w_{V}} X]  \label{eq:vform2}
\end{equation}%
is a bright soliton in the parameter region $\frac{g_{0}+g_{2}}{g_{V}}\leq
\frac{\omega _{10}}{2\omega _{V}}$ and $w_{V}<0$, and $U$ is a dark soliton.
Such dark-bright solution at two band minima is different from previous
bright-bright or dark-dark soliton solutions in a spin-1/2 BEC \cite%
{stripeso,Wu,Xu,Achilleos}. The stripe solitons for spin states $\left\vert
\uparrow \right\rangle $ and $\left\vert \downarrow \right\rangle $ are
formed by the superposition of such bright-dark solitons at two band minima,
leading to strong out-of-phase density modulations for two states, as shown
in Fig.~\ref{Fig1}(b). The total density $n\equiv 2|U|^{2}+|V|^{2}=\frac{%
-2\omega _{10}}{2g_{0}+g_{2}}-\frac{g_{2}}{2g_{0}+g_{2}}|V|^{2}$ possesses a
weak soliton profile ($\sim |V|^{2}$) on top of a uniform background $\frac{%
-2\omega _{10}}{2g_{0}+g_{2}}$ [see Fig.~\ref{Fig1}(b)]. Similar types of
solitons also exist for $\Omega >1$ although analytic solutions are hard to
obtain.

\begin{figure}[tb]
\begin{center}
\includegraphics[width=1.0\linewidth]{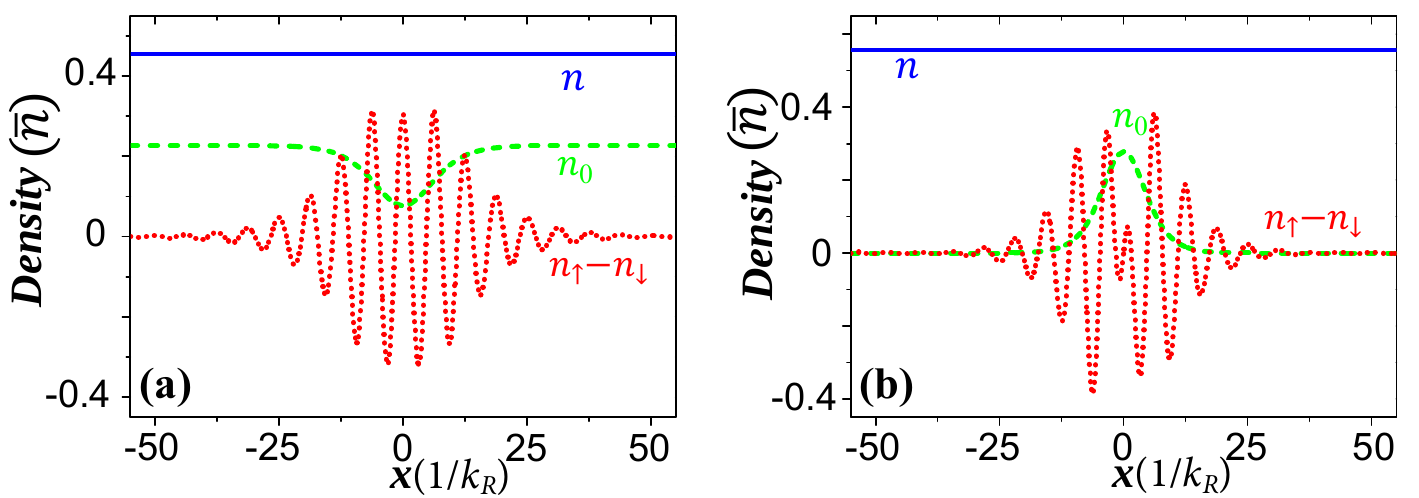}
\end{center}
\caption{(a) and (b) The spin-density distributions for the magnetic stripe
soliton with $g_{2}>0$ and $g_{2}<0$, respectively. The total density is
uniform, while the spin density $\mathcal{F}_{z}=n_{\uparrow }-n_{\downarrow
}$ exhibits stripe modulation. The parameters are $g_{2}=0.1$, $\protect%
\omega _{20}=-48.5$ for (a) and $g_{2}=-0.1$, $\protect\omega _{20}=-\frac{%
g_{0}}{g_{0}+g_{2}}\protect\omega _{10}$ for (b) with other parameters $%
\protect\epsilon =0.1$, $g_{0}=1$, $\protect\gamma =2g_{2}$, $\Omega =1$, $%
\protect\omega _{10}=-50$, $\protect\delta =0.5$. }
\label{Fig2}
\end{figure}

\emph{Magnetic stripe solitons}--- For typical nonlinear interactions in Eq.~%
\ref{eq:eom} with $\gamma=0$, the dark soliton dip cannot be perfectly filled by the bright
soliton particles, leading to a small dip for the total density. We note that it is possible to construct the
magnetic stripe soliton with the uniform total density, with
$\gamma\neq 0$.

For $\Omega =1$, the soliton solution is still static and $V$ takes the same
form (\ref{eq:vform1}) or (\ref{eq:vform2}) as that for $\gamma =0$, except
that $g_{V}$ and $\omega _{V}$ change to $g_{V}=\frac{%
2g_{2}^{2}+3g_{0}g_{2}-g_{0}\gamma /2}{2g_{0}+g_{2}+\frac{\gamma }{2}}$, $%
w_{V}=\frac{2(g_{0}+g_{2})\omega _{10}}{2g_{0}+g_{2}+\frac{\gamma }{2}}%
-\omega _{20}$. Eq. (\ref{eq:uv}) for determining $U$ is modified by
replacing $2g_{0}+g_{2}$ in the denominator with $2g_{0}+g_{2}+\gamma /2$.
The total density is $n=\frac{-2\omega _{10}}{2g_{0}+g_{2}+\gamma /2}-\frac{%
g_{2}-\gamma /2}{2g_{0}+g_{2}+\gamma /2}|V|^{2}$, which becomes uniform when
$\gamma =2g_{2}$. The spin-density profiles for the magnetic stripe soliton
with (anti-)ferromagnetic interaction $g_{2}<0$ ($g_{2}>0$) are shown in
Figs. 2(a) and 2(b). While the total density is uniform, the spin density $%
\mathcal{F}_{z}=n_{\uparrow }-n_{\downarrow }$ exists stripe density
modulation. For $g_{2}<0$, the magnetic stripe soliton may be formed either
by dark and bright solitons or by dark and anti-dark solitons, while the
latter may have striped spin background (see Appendix).

\begin{figure}[tbh]
\begin{center}
\includegraphics[width=1.0\linewidth]{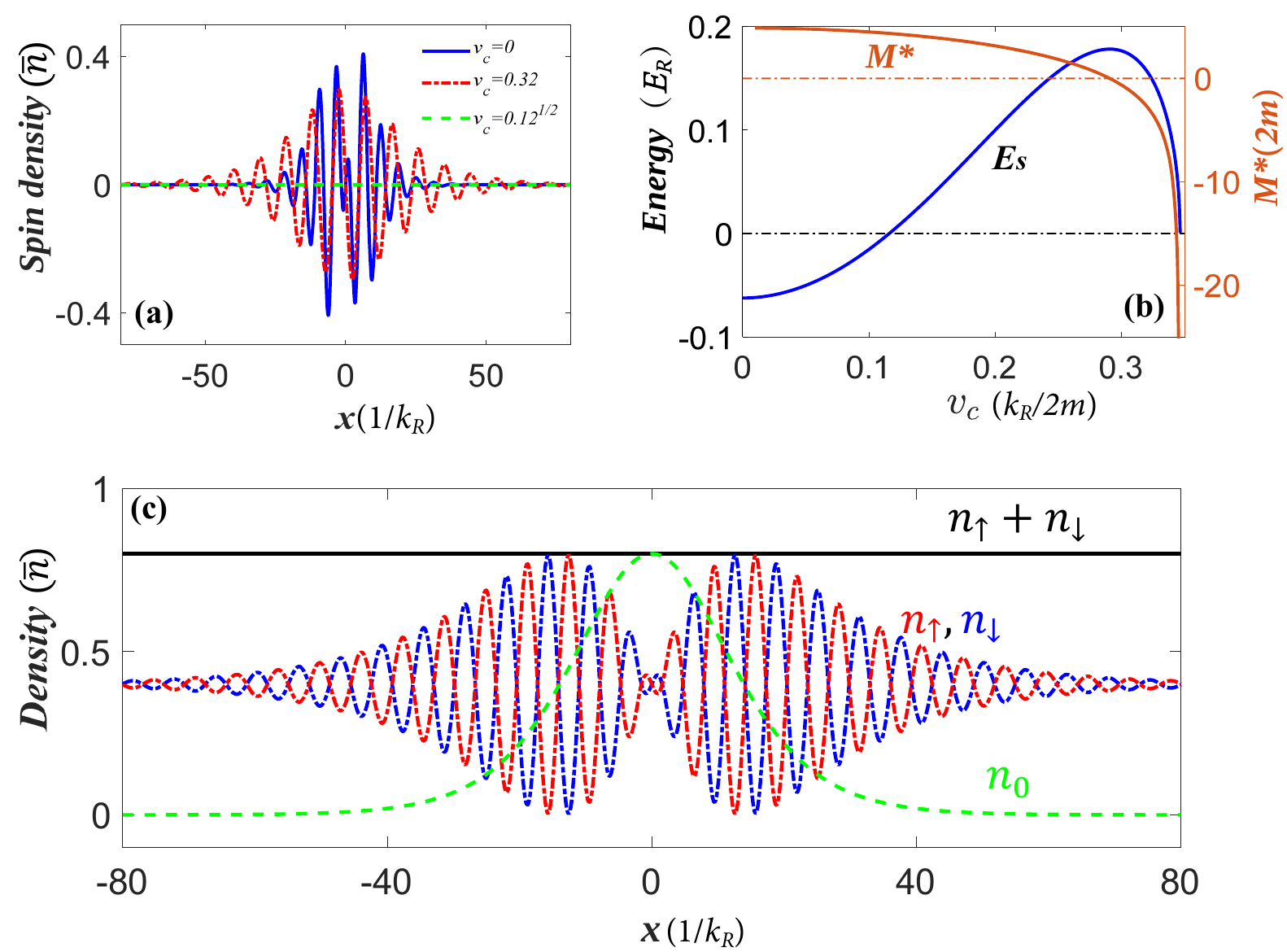}
\end{center}
\caption{(a) The moving magnetic stripe soliton profiles at different
velocities. (b) The energy $E_{s}$ (blue line) and effective mass $M^{\ast }=%
\frac{\partial E_{s}}{\partial v_{c}^{2}}$ (red line) of the magnetic stripe
soliton versus its velocity. Other parameters for (a) and (b) are $\protect%
\epsilon =0.1$, $g_{2}=-0.1$, $g_{0}=1$, $\Omega =2$, $\protect\gamma %
=4g_{2} $, $C=60$, $\protect\delta =1.5055$. (c) Density profiles for
localized stripe waves. The spin tensor density $\mathcal{F}%
_{z}^{2}=n_{\uparrow }+n_{\downarrow }$ is a constant. The parameters are $%
\protect\epsilon =0.1$, $g_{2}=-0.1$, $g_{0}=0.19$, $\Omega =1$, $\protect%
\gamma =-2g_{0}$, $\protect\omega _{10}=-7.2$, $\protect\omega _{20}=\frac{%
g_{0}}{g_{0}+g_{2}}\protect\omega _{10}$, $\protect\delta =0.5$.}
\label{Fig3}
\end{figure}

For $\Omega >1$, the general soliton solution for an arbitrary $\gamma $ is
not easy to obtain. However, the magnetic stripe soliton with a constant
total density $n=C$, which requires $\gamma =g_{2}(6-\frac{4}{\Omega })$,
can be obtained analytically. Such magnetic stripe soliton for $\Omega >1$
can have a finite velocity, with the wave function%
\begin{eqnarray}
U(X,T) &=&\sqrt{\frac{n_{v}}{2}}\sech [f(X- v T)]\exp (i\delta kX), \\
V(X,T) &=&\sqrt{n_{v}}\tanh [f(X-vT)]+\frac{iv}{\sqrt{2C|g_{2}|}}
\end{eqnarray}%
for $g_{2}<0$. Here $f(X-vT)=\sqrt{|g_{2}|n_{v}}(X-vT)$, $n_{v}=C+\frac{v^{2}%
}{2g_{2}}$, and $\delta k=\frac{\Omega v}{2(\Omega -1)}$ corresponds to a
small derivation of the center momentum away from the band minimum. Notice
that the solution is valid only when $\Omega $ is not so close to 1 to
ensure a small $\delta k$. $T=\epsilon ^{2}t$ describes the slowly varying
time, and the velocity parameter $v\leq \sqrt{2C|g_{2}|}$ should be always
less than the sound speed of the background. The soliton energies induced by
the nonlinear interaction read $\omega _{10}=-g_{0}C-g_{2}C\frac{3\Omega -1}{%
2\Omega }-v^{2}\frac{2\Omega ^{2}-2\Omega +1}{4\Omega (\Omega -1)}$, and $%
\omega _{20}=-g_{0}C$. Such magnetic stripe soliton has a velocity $%
v_{c}=\epsilon v$ with respect to the static particle density background.
The profile of the magnetic stripe soliton depends on its velocity: the peak
value decreases while its spatial width increases with the velocity [as
shown in Fig. 3 (a)]. Similar soliton solutions can be obtained for $g_{2}>0$%
.

The moving soliton usually can be described as an quasiparticle with an
effective mass and energy. The energy $E_{s}$ of the soliton is defined as
the difference between the grand canonical energies in the presence and
absence of the soliton (with the same particle density background), while
the effective mass $M^{\ast }=\frac{\partial E_{s}}{\partial (v_{c}^{2})}$
\cite{Qu1,Scott} is determined accordingly. In Fig. 3 (b), the energy and
effective mass of the moving stripe soliton are plotted versus its velocity.
The energy increases first and then decreases as the velocity increases,
while the effective mass possesses both negative and positive values and
crosses zero at a velocity smaller than the sound speed. Close to the sound
speed, the effective mass approaches infinity due to the velocity dependence
of the dark-bright soliton profiles. These characters, especially the zero
mass at a velocity less than the sound speed, are in sharp contrast to
bright or dark solitons with a fixed sign of mass in previous studies \cite%
{eff1,eff2,eff3,eff4,Scott}. Recently, the positive mass and negative mass of soliton was discussed in a spin-orbit coupled BEC~\cite{stripeso2}, which is induced by the effective mass of single atom  (single-particle band dispersion).
The negative mass of magnetic stripe soliton is mainly induced by the
nonlinearity instead of the single-particle band dispersion. Therefore, the
magnetic stripe soliton can admit negative mass even though the effective
single-particle mass is positive, leading to very unique properties such as
the infinity and zero mass at different velocities.
We note that similar tunable inertial mass of atomic Josephson vortices and related solitary waves can also be achieved through an adjustable linear coupling between the two components in a BEC \cite{Brand}. These striking inertial mass properties could inspire more discussions on the dispersion relation of quasi-particle nonlinear waves.


\begin{figure}[tb]
\begin{center}
\includegraphics[width=1.0\linewidth]{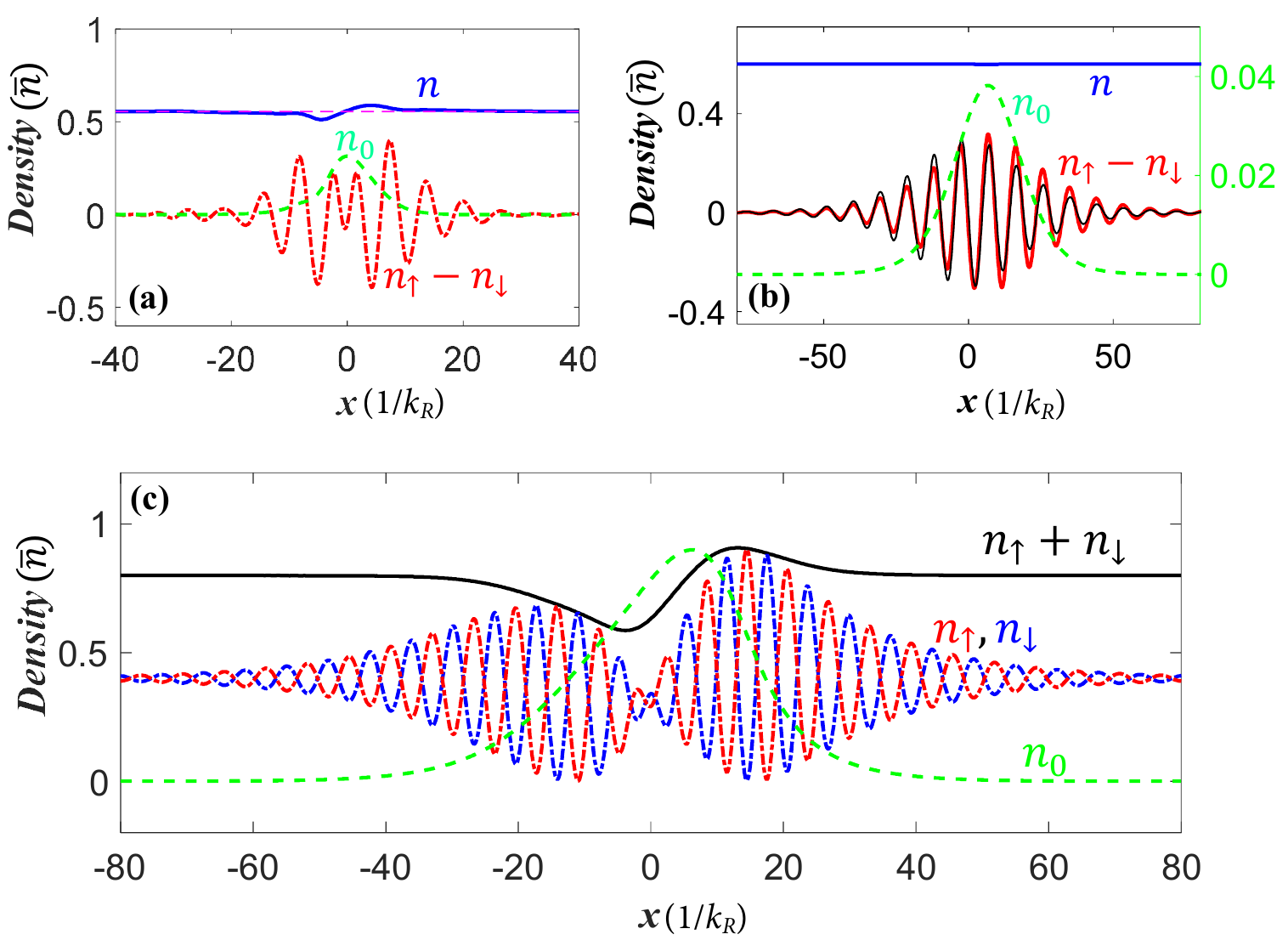}
\end{center}
\caption{(a) and (b) Numerical simulations for the magnetic stripe solitons
in Fig.~2(b) and Fig.~3(b). (c) Numerical simulation for the localized
stripe wave in Fig.~3(c). The evolution time is $t=20$. The solitons are
stable and the high order corrections are negligible in (a) and (b) but more
significant for (c). The thin purple dashed line in (a) is the initial total
density, and the thin black solid line in (b) is the initial stripe density
modulation. }
\label{Fig4}
\end{figure}

\emph{Localized stripe wave}--- In stripe solitons discussed in either
spin-1/2 BEC \cite{stripeso,Wu,Xu,Achilleos,stripeso2} or above magnetic
stripe soliton, the striped density modulation resides on a bright or dark
soliton background. Here such density background is given by $|U|^{2}+|V|^{2}
$, which can be tuned to be uniform by the spin-dependent interaction $%
\gamma $, therefore the stripe soliton for states $\left\vert \uparrow
\right\rangle \ $and $\left\vert \downarrow \right\rangle $ is neither
bright nor dark, but a localized stripe wave. For instance, when $\Omega =1$%
, $\gamma =-2g_{0} $, the spin tensor density $\mathcal{F}%
_{z}^{2}=n_{\uparrow }+n_{\downarrow } $ becomes uniform. For $g_{2}>0$ ($%
g_{2}<0$), the localized stripe waves for states $\left\vert \uparrow
\right\rangle \ $and $\left\vert \downarrow \right\rangle $ exists for $%
\frac{2w_{V}}{g_{V}}\leq \frac{-\omega _{10}}{g_{0}+g_{2}}$ ($\frac{w_{V}}{%
|g_{V}|}=\frac{-\omega _{10}}{g_{0}+g_{2}}$), while the state $|0\rangle $
admits a dark (bright) soliton profile, as shown in Fig.~3(c).

\emph{Stability of solitons}--- To check the stability of these stripe
solitons obtained from the multi-scale expansion method, we numerically
simulate the time evolution dynamics of the obtained solitons using the GP
equation (\ref{eq:eom}) with the analytic soliton solutions as the initial
states. As an example, we consider the magnetic stripe solitons for
ferromagnetic interaction ($g_{2}<0$). Both static [$\Omega =1$ in Fig.
2(b)] and moving [$\Omega >1$ and $v_{c}=0.32$ in Fig. 3(a)] solitons are
found to be stable [see Fig.~4(a) and (b)]. We notice that higher order
terms in the solution may induce slight modulation of the soliton profile,
which is typically of the order of $\epsilon ^{2}$. In addition, a nonzero $%
\gamma $ breaks the balance between state $|+\rangle $ and $|0\rangle $,
leading to small atom transitions between them. However, as long as $\gamma $
is weak compared with the density interaction $g_{0}$, such transitions
hardly affect the stability of the soliton and the uniformity of the total
density.

The localized stripe wave requires a $\gamma $ that is comparable with $%
g_{0} $, yielding a strong coupling between states $|+\rangle $ and $%
|0\rangle $. Nevertheless, we find that stable solitons can still exist when
both $\gamma $ and $g_{0}$ are weak compared with the recoil energy. In
Fig.~4(c), we show our numerical result with density interaction $%
g_{0}=1.9|g_{2}|$, $g_{2}<0$ and $\gamma =-2g_{0}$. Although the localized
stripe wave is stable, the spin background $|U|^{2}+|V|^{2}$ is no longer
uniform, but exhibits a soliton profile due to the nonzero $\gamma $ (which
is comparable with $g_{0}$). The numerical simulations of the dark-bright
stripe solitons with $\gamma =0$ are shown in the Appendix, which are in
good agreement with Fig.~1(b) thanks to the absence of transitions between
states $|+\rangle $ and $|0\rangle $.

\emph{Discussion and conclusion--- }The spin interaction $g_{2}$ (either $>0$
or $<0$) of alkaline atoms is usually much smaller than the density
interaction $g_{0}$~\cite{PhysRevLett.81.742, Ohmi1998Bose,
RevModPhys.85.1191}. The intra-component
interaction strength $\gamma$ for the state $\left\vert
0\right\rangle $, can be realized by using an additional Feshbach resonance~\cite{RevModPhys.82.1225}.
In our simulation of the magnetic stripe soliton, we
consider $g_{2}=\pm 0.1g_{0}$, and as we increase (decrease) $|g_{2}|$, the
soliton properties would not be affected, except that its size may decrease
(increase) slightly. Different from the ground-state stripe phase, the
stripe solitons can exit in a large parameter region and the choice of $%
\Omega $ and $\Delta $ is very flexible. This is because the solitons are
metastable states and we can have solutions even when the detuning between
two (bright and dark) band minima are much larger than $g_{2}$. We focus our
study in the $\Omega \geq 1$ region. For $\Omega <1$, the bright band has
two band minima around $k=1$, and there still exist magnetic stripe soliton
solutions with $k_{1}$ located at one of two band minima (see Appendix). In
experiments, the soliton may be imprinted at the center of the BEC cloud
using light-induced spatial dependent potential, and observed after some
evolution time~\cite{S8,Denschlag,Anderson}.

In summary, we demonstrate new types of stripe solitons, such as magnetic
stripe solitons and localized stripe waves, can be engineered by tuning the
band dispersion in a spin-1 BEC with spin-orbit coupling. These solitons are
quite different from the solitons in the spin-orbit coupled BEC~\cite%
{stripeso,Wu,Achilleos,stripeso2,Kartashov}, which further enrich soliton
family in BEC and many other nonlinear systems and may be used for both
fundamental studies (e.g., spin-orbital coupling, nonlinear effects, quantum
fluctuations~\cite{Kadau,QF,QF2}, modulational instability~\cite{MI},
quantum entanglement~\cite{Gertjerenken}) and realistic applications (e.g.,
atomic soliton laser). The striking effective mass characters of these
solitons may shed light on the study of particle physics \cite{MRN,PAR}.
While many interesting problems, such as the generalization of magnetic
stripe solitons to higher dimension \cite{Zhou,Gautam,Gallem,Law}, remain to
be explored, our work clearly showcases the power of soliton generation
through band dispersion engineering, which may provide a new approach for
exploring soliton physics in many physical branches.

\begin{acknowledgments}
LCZ is supported by National
Natural Science Foundation of China (Contact No. 11775176), Major Basic
Research Program of Natural Science of Shaanxi Province (Grant No.
2018KJXX-094), China Scholarship Council. XWL and CZ are supported by Air
Force Office of Scientific Research (FA9550-16-1-0387), National Science
Foundation (PHY-1806227), and Army Research Office (W911NF-17-1-0128).
\end{acknowledgments}

\newpage

\begin{widetext}
\setcounter{figure}{0} \renewcommand{\thefigure}{A\arabic{figure}} %
\setcounter{equation}{0} \renewcommand{\theequation}{A\arabic{equation}}

\section{Appendix}
In the appendix, we derive the non-linear Schr\"{o}dinger
equation of the spin-1 system, and give the magnetic stripe soliton
solutions in other parameter regions not shown in the main text.

\subsubsection{Non-linear Schr\"{o}dinger equation}

In the Lab frame, the second-quantization Hamiltonian can be written as
\begin{equation*}
\mathcal{H}=\int dx\Psi ^{\dag }H_{0}\Psi +\int dx\frac{g_{0}}{2}(\Psi
^{\dag}\Psi )^{2}+\frac{\gamma} {2} (\psi _{0}^{\dag} \psi _{0})^{2}+\frac{g_{2}}{2}(\Psi ^{\dag}\mathbf{F}\Psi )^{2},
\end{equation*}%
with spin operator $\mathbf{F}=(F_{x},F_{y},F_{z})$, and atom field $\Psi
=(\psi _{\uparrow },\psi _{0},\psi _{\downarrow })$. The non-linear Schr\"{o}%
dinger equation can be obtained by
\begin{equation*}
i\partial _{t}\Psi =[\Psi ,\mathcal{H}].
\end{equation*}%
In the quasi-momentum frame, we have~\cite{Luo}
\begin{equation*}
i\partial _{t}\psi _{j}=H_{0}\psi _{j}+(g_{0}\bar{n}+g_{2}\bar{n})n\psi
_{j}-g_{2}|\psi _{j}|^{2}\psi _{j}+\delta _{j,0}\gamma |\psi _{j}|^{2}\psi _{j}+g_{2}\psi _{j}^{\ast }Q_{j}(x),
\end{equation*}%
Where $j=\pm ,0$ and $Q_{+}(x)=\psi _{-}^{2}+\psi _{0}^{2}e^{i4x}$, $%
Q_{-}(x)=\psi _{+}^{2}-\psi _{0}^{2}e^{i4x}$, $Q_{0}(x)=(\psi _{+}^{2}-\psi
_{-}^{2})e^{-i4x}$. We are interested in the solutions with momenta
centering at the band minima, while the last term involves couplings with
higher momenta far away from the band minima, therefore its effects is
negligible and can be omitted. This is also confirmed by our numerical
simulation in Fig.~\ref{FigS1}, where the last term only induces tiny and
fast spatial modulations without affecting the soliton profile.

\begin{figure}[b]
\begin{center}
\includegraphics[width=0.7\linewidth]{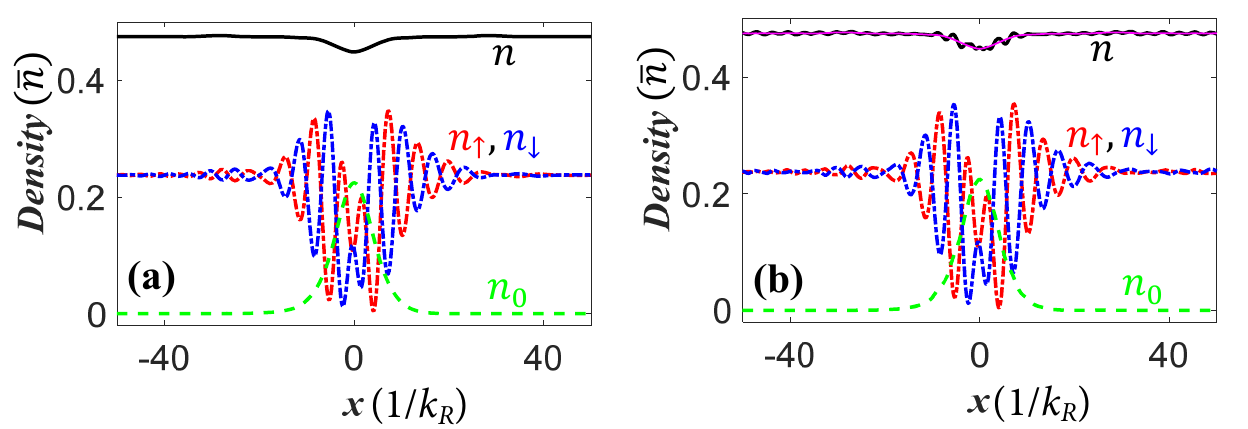}
\end{center}
\caption{(a) and (b) Numerical results of the soliton profiles at $t=20$
without and with the $Q_{j}$ terms. The parameters are the same as those in
Fig.~1b in the main text. The thin purple line in (b) is the total density
without $Q_{j}$ term [same as the black line in (a)].}
\label{FigS1}
\end{figure}

\subsubsection{Dark-anti-dark magnetic stripe solitons}

In the main text, we have focused on magnetic stripe solitons formed by dark
and bright solitons, which admit spin-balanced background. However, at $%
\Omega =1$, the spin background can be imbalanced if it is formed by dark
and anti-dark solitons. Such dark and anti-dark solitons can exist for
ferromagnetic interactions $g_{2}<0$ with proper choice of parameters (i.e.,
$\omega _{10}$ and $\omega _{20}$),
as shown in Fig.~\ref{FigS2} (its stability is confirmed numerically). The
soliton resides on a striped spin background, which is different from the
stripe magnetic solitons with a zero spin background (as shown in Fig.~2 of
the main text).

\begin{figure}[tb]
\begin{center}
\includegraphics[width=0.4\linewidth]{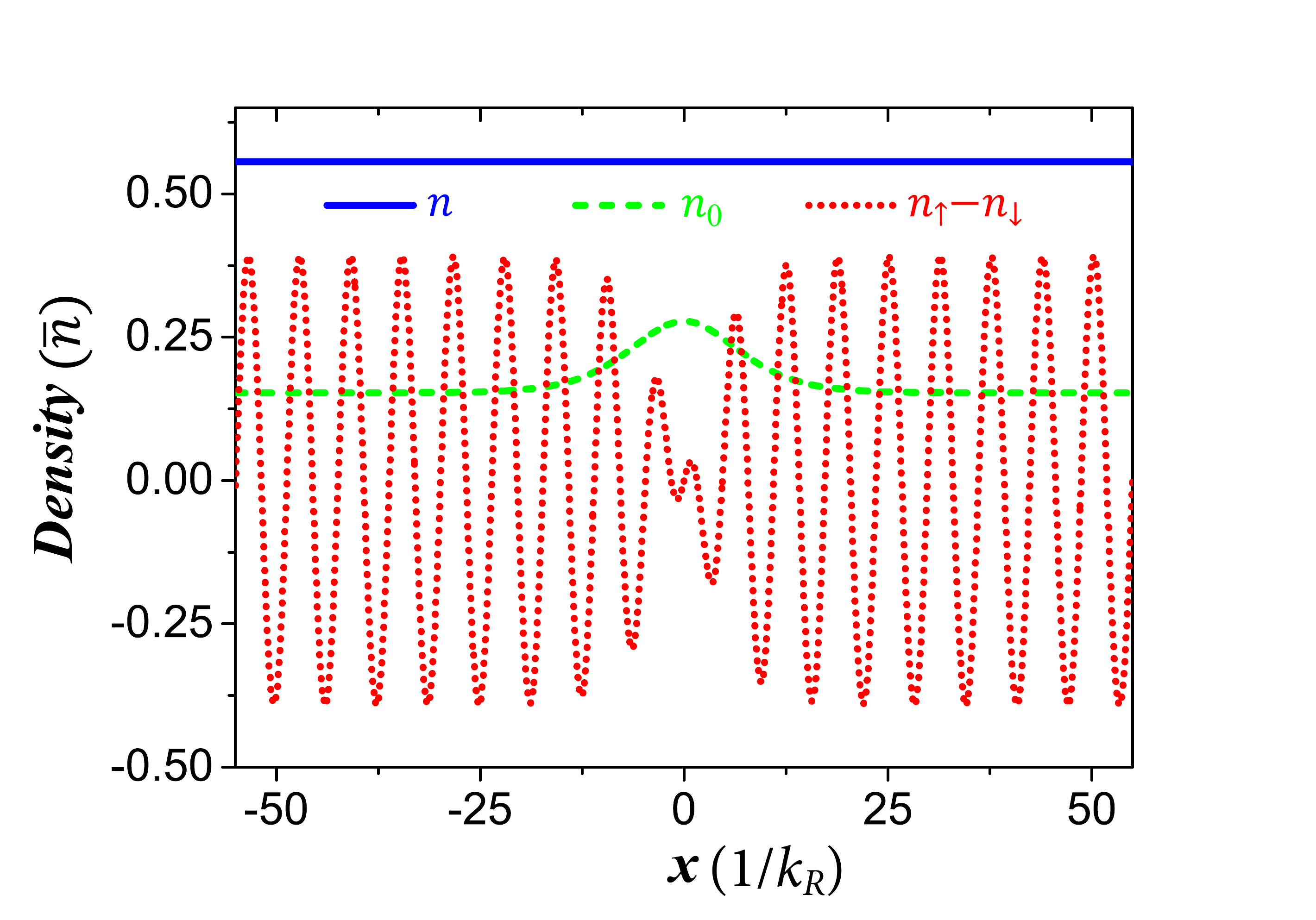}
\end{center}
\caption{The spatial profiles of dark-anti-dark magnetic stripe soliton. The
parameters are $\protect\epsilon =0.1$, $g_{2}=-0.1$, $g_{0}=1$, $\protect%
\omega _{10}=-50$, $\protect\omega _{20}=-52.5$, $\protect\delta =0.5$. }
\label{FigS2}
\end{figure}

\subsubsection{Magnetic stripe solitons for $\Omega<1$}


For $\Omega <1$, the bright band has two minima at $k=1\pm \sqrt{1-\Omega
^{2}}$, thus we expect to find stable stripe solitons by choosing the center
momentum as $k_{1}=1\pm \sqrt{1-\Omega ^{2}}$. 
Magnetic stripe solitons with a uniform total density require $\gamma =\frac{%
2g_{2}(3\Omega ^{2}-4)(\Omega ^{2}+2\sqrt{1-\Omega ^{2}}-2)}{\Omega ^{4}}$.
As an example, we present a stripe soliton solution for $g_{2}<0$ and $%
k_{1}=1-\sqrt{1-\Omega ^{2}}$, while similar soliton solutions can be
obtained in other parameter regimes. The stripe magnetic soliton
wavefunctions are $\psi _{\uparrow }\approx \frac{1}{\sqrt{2}}\epsilon
\lbrack \frac{\sqrt{1-\Omega ^{2}}-1}{\Omega }U(X,T)e^{ik_{1}x-i\omega
_{1}t}+V(X,T)e^{ik_{2}x-i\omega _{2}t}]$ , $\psi _{0}\approx \epsilon
U(X,T)e^{ik_{1}x-i\omega _{1}t}$, and $\psi _{\downarrow }\approx \frac{1}{%
\sqrt{2}}\epsilon \lbrack \frac{\sqrt{1-\Omega ^{2}}-1}{\Omega }%
U(X,T)e^{ik_{1}x-i\omega _{1}t}-V(X,T)e^{ik_{2}x-i\omega _{2}t}]$, where $%
U(X,T)$ and $V(X,T)$ are
\begin{eqnarray}
U(X,T) &=&\sqrt{\frac{p}{\gamma _{r}}}\sech[\sqrt{\frac{p}{2}} (X-vT)]\exp
\left( -\frac{iT_{1}[n_{1}(g_{0}+g_{2})+\omega _{10}]}{1-\Omega ^{2}}+\frac{%
ipT_{1}}{2}-\frac{1}{4}iT_{1}v_{1}^{2}+\frac{iv_{1}X}{2}\right) , \\
V(X,T) &=&\sqrt{n_{1}}[\sqrt{1-\frac{v^{2}}{2(-g_{2})n_{1}}}\tanh [\sqrt{%
\frac{1}{2}(-g_{2})n_{1}}\sqrt{1-\frac{v^{2}}{2(-g_{2})n_{1}}}(X-vT)]+\frac{%
iv}{\sqrt{2(-g_{2})n_{1}}}]\exp [i\phi ],
\end{eqnarray}%
with $\gamma _{r}=\frac{\frac{g_{2}(2-\Omega ^{2})(1-\sqrt{1-\Omega ^{2}})}{%
\Omega ^{2}}-\frac{1}{2}\gamma (\sqrt{1-\Omega ^{2}}+1)}{1-\Omega ^{2}}$, $p=%
\frac{1}{2}(-2g_{2}n_{1}-v^{2})$, $v_{1}=\frac{v}{1-\Omega ^{2}}$, $%
T_{1}=(1-\Omega ^{2})T$, and $\phi =g_{2}n_{1}T-n_{1}(g_{0}+g_{2})T+\omega
_{20}T$.

\begin{figure}[tb]
\begin{center}
\includegraphics[width=0.4\linewidth]{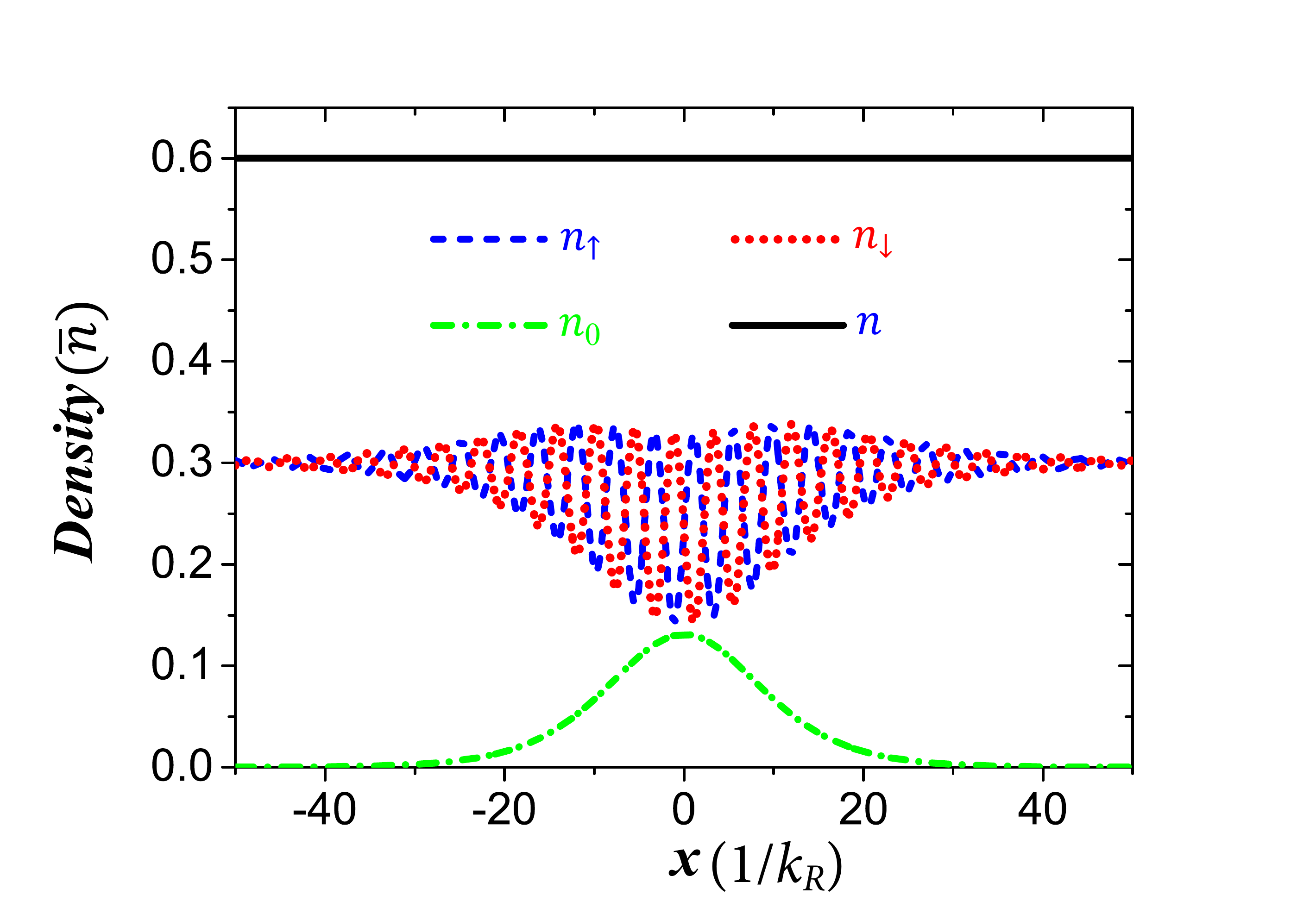}
\end{center}
\caption{The magnetic stripe soliton for $\Omega <1$ and $g_{2}<0$.
Numerical test shows that it is stable. The center momentum is $k_{1}=1-%
\protect\sqrt{1-\Omega ^{2}}$, and the additional spin-dependent modulation
coefficient is $\protect\gamma =\frac{2g_{2}(3\Omega ^{2}-4)(\Omega ^{2}+2%
\protect\sqrt{1-\Omega ^{2}}-2)}{\Omega ^{4}}$. Other parameters are $%
\protect\epsilon =0.1$, $g_{2}=-0.1$, $g_{0}=1$, $\Omega =2/3$, $n_{1}=60$, $%
\protect\omega _{10}=0$, $\protect\omega _{20}=0$, $\protect\delta =4/18$, $%
v=3$. }
\label{FigS3}
\end{figure}
Fig.~\ref{FigS3} shows the typical density (spin) profiles of such solitons.
The soliton is confirmed to be stable in the GP equation simulation,
although higher order terms in the solution may induce minor distortion.
\end{widetext}

\end{document}